\documentclass[12pt, preprint]{aastex}

\usepackage {epsf}
\usepackage {lscape, graphicx}
\usepackage {psfig}
\usepackage {amssymb,amsmath}
\usepackage [figuresleft]{rotating}
\usepackage{amsfonts}
\usepackage{natbib}
\usepackage{subfigure}

\def\deg{\mbox{$^\circ$}}

\newcommand{\sersic}{S\'{e}rsic }

\begin{document}

\title{The CFHT Legacy Survey: \\
The Morphology-Density Relation of Galaxies out to $z\sim1$\altaffilmark{1}}
\author{Martijn J. H. M. Nuijten\altaffilmark{2,3,5}, Luc 
Simard\altaffilmark{3}, Stephen Gwyn\altaffilmark{4,5}, Huub J. A. 
R\"{o}ttgering\altaffilmark{2}}
\altaffiltext{1}{This work is based in part on data obtained as part of the 
Canada-France-Hawaii  Telescope Legacy Survey, a collaborative project of the 
National Research  Council of Canada and the French Centre national de la 
recherche scientifique.}
\altaffiltext{2}{Leiden Observatory, P.O. Box 9513, NL-2300 RA, Leiden, The
Netherlands.}
\altaffiltext{3}{National Research Council of Canada, Herzberg Institute of 
Astrophysics, 5071 West Saanich Road, Victoria, BC V9E 2E7, Canada.}
\altaffiltext{4}{Department of Physics \& Astronomy, University of Victoria, 
P.O. Box 3055, Victoria, BC V8W 3P6, Canada.}
\altaffiltext{5}{Guest User, Canadian Astronomy Data Centre, which is 
operated by the Dominion Astrophysical Observatory for the National Research 
Council of Canada's Herzberg Institute of Astrophysics.}

\begin{abstract}
We study the relationships between galaxy total luminosity ($M_{g'}$),
morphology, color and environment as a function of redshift. We use a
magnitude-limited sample of 65,624 galaxies in the redshift range
$0<z<1.3$ taken from one of the $1\deg\times1\deg$
Canada--France--Hawaii Telescope Legacy Survey Deep Fields. We
parametrize galaxy morphology according to the S\'ersic index $n$, 
taking $n > 2$ to be ``bulge-dominated'' and  $n < 2$ to be ``disk-dominated''. 
Our $n > 2$ number fractions at $z=0.1$ agree well with those based on 
Sloan Digital Sky Survey galaxies. We find that the $n > 2$ galaxy number 
fraction is constant with redshift in the field. However, for overdense 
environments this fraction is larger and increases towards lower redshifts, 
higher densities and higher luminosities. Rest-frame color-magnitude 
diagrams show that the color distribution is bimodal out to our redshift 
limit of $z\sim1$ with a prominent red-sequence of galaxies at $0.2<z<0.4$ 
and a large blue-peak dominance at $0.8<z<1$. We use this bimodality to 
define a red fraction as the fraction of galaxies having a rest-frame color 
$u^{*}-g'>1$. For all environments, this fraction increases towards lower 
redshifts and higher luminosities. The red fraction within cluster-like 
regions changes $60\%$ faster with redshift as compared to the field for 
galaxies with $M_{g'}<-19.5$. Using, for the first time, observations across 
many cluster-field interfaces distributed over a single, large volume, we 
trace the large-scale morphology-density relation and the Butcher-Oemler 
effect over a period of almost $8$ Gyr.

\end{abstract}


\section{Introduction}

Galaxy populations show great diversity in morphologies and colors
\citep[e.g.,][]{hubble26}. Many studies of the local and high-redshift
universe show that there exist strong relations between galaxy morphology, 
color and environment 
\citep[e.g.,][]{dressler80,kennicutt83,whitmore93,dressler97,hogg03}.

In order to explain the observed correlations, current theories of
galaxy formation and evolution adopt a $\Lambda$-dominated cold dark
matter model for the universe where large-scale structure is built 
through hierarchical clustering of dark matter halos. Early-type 
galaxies are then assembled through the (successive) merging of disks 
that formed in these halo's. An important test of this formation scenario 
is to trace and quantify the redshift evolution of observable
galaxy properties as a function of environment. Significant progress
has recently been made on this issue with the availability of large
and homogeneous datasets for direct comparisons with theoretical
predictions. For example, the Sloan Digital Sky Survey
\citep[SDSS;][]{stoughton02} has been used at the low redshift end
($z\sim0.1$) to study various correlations between observable galaxy
properties and the larger scale environment \citep{goto03,hogg04}.
At higher redshifts, \citet{bell04} were able to put strong constraints  
on hierarchical formation models by exploring the rest-frame properties 
of color-selected early-type galaxies as function of redshift. In this 
Letter, we include galaxy environment as another important variable 
in the evolutionary signatures predicted by the hierarchical models. 
Taking the SDSS as baseline \citep{hogg04} we examine both the morphology 
(bulge- vs. disk-dominated) and rest-frame color (red vs. blue) redshift 
evolution of galaxies residing within different environments from the 
field to massive clusters. We make use of data from the 
Canada--France--Hawaii Telecope Legacy Survey (CFHTLS) which is well 
suited to address these issues as its extensive sky coverage 
provides large, homogeneous and statistically well-defined samples of 
galaxies spanning a wide range of redshifts and densities.

This Letter is organized as follows: Section~\ref{data} describes the
CFHTLS-Deep data, and Section~\ref{analysis} outlines the morphology
and environment analysis. In Section~\ref{results}, we present and
discuss our results. The cosmological parameters adopted throughout
this Letter are $H_0=70\,\mathrm{km\,s^{-1}\,Mpc^{-1}}$, and
$(\Omega_\mathrm{M},\Omega_\mathrm{\Lambda},\Omega_\mathrm{k})=(0.3,0.7,0.0)$.

\section{The Data}\label{data}

The CFHTLS-Deep survey consists of $u^{*}g'r'i'z'$ images of four
uncorrelated areas of the sky taken with the MegaPrime prime focus
mosaic imager (thirty-six 2080x4622 chips). Each patch covers about
$1\deg\times1\deg$ and is located away from the galactic plane in
order to minimize galactic extinction and bright star
contamination. When completed, the Deep Survey will produce images
almost as deep as the Hubble Deep Field \citep{williams96} but
covering an area of the sky 3500 times as large. The data used here are
from the CFHTLS-D3 field (RA(2000)$\,=14^{h}\,19^{m}\,28^{s}$,
DEC(2000)$\,=+52^{o}\,40'\,41''$) after one year of observing. D3 was
chosen because it overlaps with a Canada-France Redshift Survey field
\citep{lilly01} and partially with the ``Groth Strip"
\citep{groth94,rhodes00}. We thus have 338 spectroscopic redshifts
available from the literature for calibrating our photometric
redshifts.

The data were processed (de-biassed, flat-fielded, fringe-corrected,
etc.) through the Elixir pipeline \citep{magnier04} and
retrieved from the Canadian Astronomical Data Centre (CADC). All the
images were astrometrically and photometrically calibrated and
then stacked (Gwyn et al., in preparation). The final photometric
zero-point uncertainties are less than $0.05$ mag. The galaxy
photometry package SExtractor \citep{bertin96} was used to detect
objects in the $i'$-band and to perform photometry using 
\citet{kron80} apertures.

\section{Analysis}\label{analysis}

\subsection{Photometric Redshifts}

Photometric redshifts and spectral types were determined 
(Gwyn et al., in preparation). Based on a comparison with
the available spectroscopic redshifts, the typical relative
photometric redshift error is $\delta z/(1+z)=0.11$. Using the
best-fit photometric redshift and spectral type, we calculated
$k$-corrections and $u^{*}g'r'i'z'$ rest-frame galaxy AB absolute
magnitudes. The maximum accessible volume $V_{acc}$ \citep{schmidt68}
was also computed for each object (Gwyn et al., in preparation) by
determining the redshift range over which its magnitude would fall
within the $i'$-band limiting magnitude of the sample ($i'<24.5$
mag). Galaxies were then weighted by the resulting $1/V_{acc}$ values
to compute number densities in the color-magnitude plane.

\subsection{Quantitative Morphologies}

We measured the structural parameters of our galaxies in the $i'$-band
in a quantitative, uniform and reproducible way using the GIM2D
package \citep{simard02}. We fit two-dimensional, PSF-convolved, pure
\sersic models of the form $\Sigma(r)=\Sigma_{0}\exp(-r/r_0)^{1/n}$ to
our galaxy images. The shape or concentration parameter $n$ is known
as the ``\sersic index'' \citep{sersic68}.  The choice of such a 
simple model was better suited to the spatial resolution of our 
ground-based data than a full, more complex bulge+disk decomposition. 
Spatially-varying point-spread-functions (FWHM$\sim0\farcs7-0\farcs9$) 
were constructed across each MegaPrime chip directly from stars in the 
images using the DAOPHOT package \citep{stetson87}. The galaxy total flux 
($F_{tot}$) and {\it intrinsic} half-light radius $r_{1/2}$ were obtained 
by integrating the best-fit GIM2D \sersic model out to $r=\infty$. 
Absolute rest-frame total $g'$-band magnitudes ($M_{g'}$) were calculated 
for the best-fit models from the apparent and absolute aperture magnitudes.

Following \citet{simard02} we conducted extensive simulations in order
to characterize the systematic biases and random errors in the
GIM2D/CFHTLS-D3 ${i'}$-band structural measurements. After careful
inspection of our simulation results, we concluded that the 
sample we will use with $i{'}_{model}<24$ mag (1) has a completeness of
$\sim98.5\%$ and (2) has combined random and systematic errors of 
$\,\bar{\Delta}n\sim{0.76}$ for $r_{1/2}>0\farcs1$.
 
\subsection{Overdensity Estimator $\delta_{15th}$}

A total of 65,624 galaxies with $i'<24.5$ mag and $0<z<1.3$ were used for 
our overdensity calculations. For each galaxy we estimate the overdensity 
of its environment relative to mean density at its redshift using the 
projected distance, $r_{15th}$, to the fifteenth nearest galaxy. The 
overdensity is then defined as the ratio of the galaxy number density 
within a cylinder of length $dD_{L}$ and radius $r_{15th}$ centered at the 
galaxy over the mean number density: 
\begin{equation}
  \delta_{15th} = (15 \,/\, \alpha\, \pi\, r_{15th}^2\, dD_{L})\, /\, 
(N_{tot}\, /\, V_{tot})\,, 
\label{odensity}
\end{equation}
\noindent where $\alpha$ is an areal correction factor for the edges of the 
MegaPrime detectors, $dD_{L}$ is the luminosity distance interval corresponding 
to a redshift interval centered at the galaxy, $N_{tot}$ and $V_{tot}$ 
are the total number of galaxies and total volume within the effective MegaPrime field 
($\sim0.9\, \mathrm{deg^2}$) of the same redshift interval.

An important issue is to properly take into account inaccuracies introduced 
by the usage of the photometric redshift. We address this by first assuming 
that each $z_{phot}$ value has an associated Gaussian error probability 
distribution. Each galaxy is then successively placed in the center of 
five adjacent redshift bins with the central bin at $z_{phot}$ and the 
outermost bins benchmarking the $99\%$ confidence level. 
We adopt $\delta z/(1+z)=0.11$ as the redshift uncertainty estimator. 
All five $\delta_{15th}$'s are weighted by the Gaussian probability distribution and 
averaged to yield one final overdensity estimate for each galaxy. Monte Carlo 
simulations show that nearly $70\%$ of the galaxies have a measurement error on 
$\delta_{15th}$ smaller than 0.3 ($68\%$ confidence level). 

\section{Results and Discussion}\label{results}

\begin{figure*}[htbp]
  \begin{center}
    \mbox{ \vspace{-1.in}
  \hspace{-.15in}
       \subfigure[\footnotesize
$0.2<z<0.4$]{\includegraphics[angle=90,scale=1.]{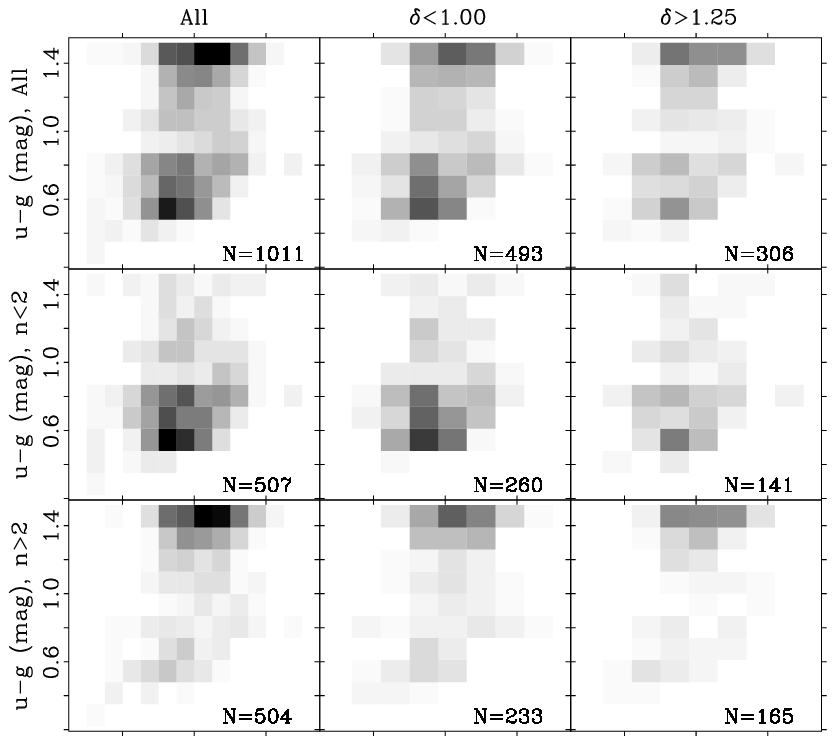}} 
\quad
      \subfigure[\footnotesize 
$0.8<z<1.0$]{\includegraphics[angle=90,scale=1.]{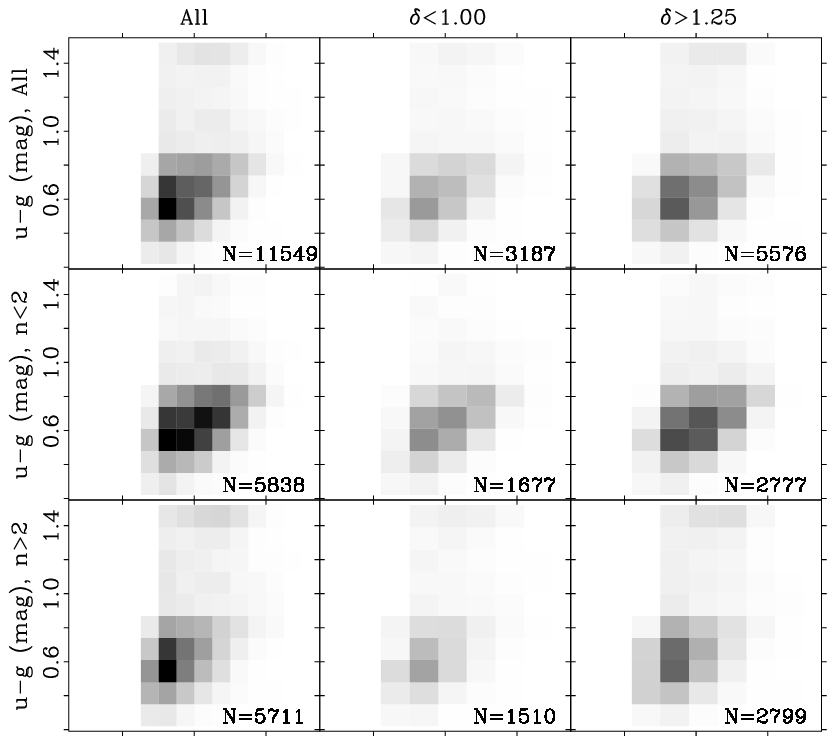}}
         }
    \caption{$1/V_{acc}$ weighted rest-frame color-magnitude diagrams 
for all galaxies with $M_{g'}<-19.5$ as function of the overdensity estimator  
$\delta_{15th}$ (rows) and \sersic index $n$ (columns) for 
(a) $0.2<z<0.4$ and (b) $0.8<z<1.0$. In each panel the grey scale 
monotonically represents the number densities of galaxies in the 
two-dimensional space of color and magnitude (for each row normalization w.r.t. 
to the highest value bin of the \emph{left} panel). Equivalent color-magnitude 
diagrams for the redshift range $0.4<z<0.8$ (not included here) show bimodal
distributions intermediate between (a) and (b).}
    \label{fig:red_seq}
  \end{center}
\end{figure*}

\begin{figure*}[htbp]
  \begin{center}
    \mbox{ \vspace{-1.5in} \hspace{-.075in}
       \subfigure[\footnotesize 
Morphology]{\includegraphics[angle=90,scale=.9]{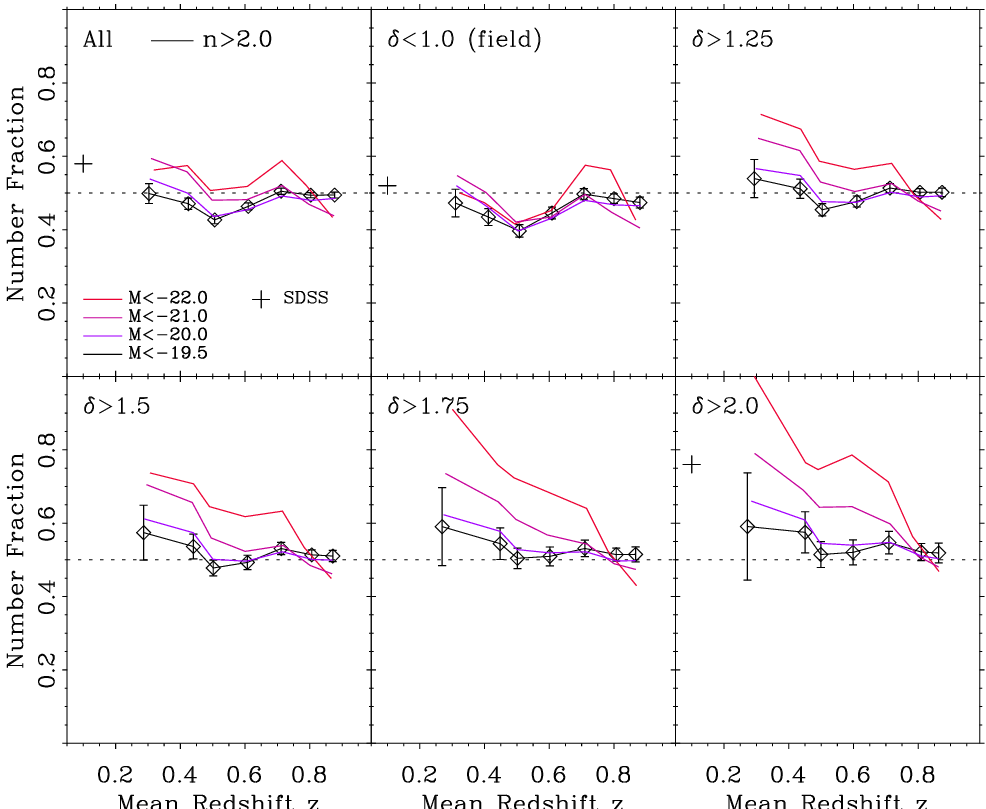}} 
\quad
 \hspace{-.25in}
       \subfigure[\footnotesize Color]{\includegraphics[angle=90,scale=.9]{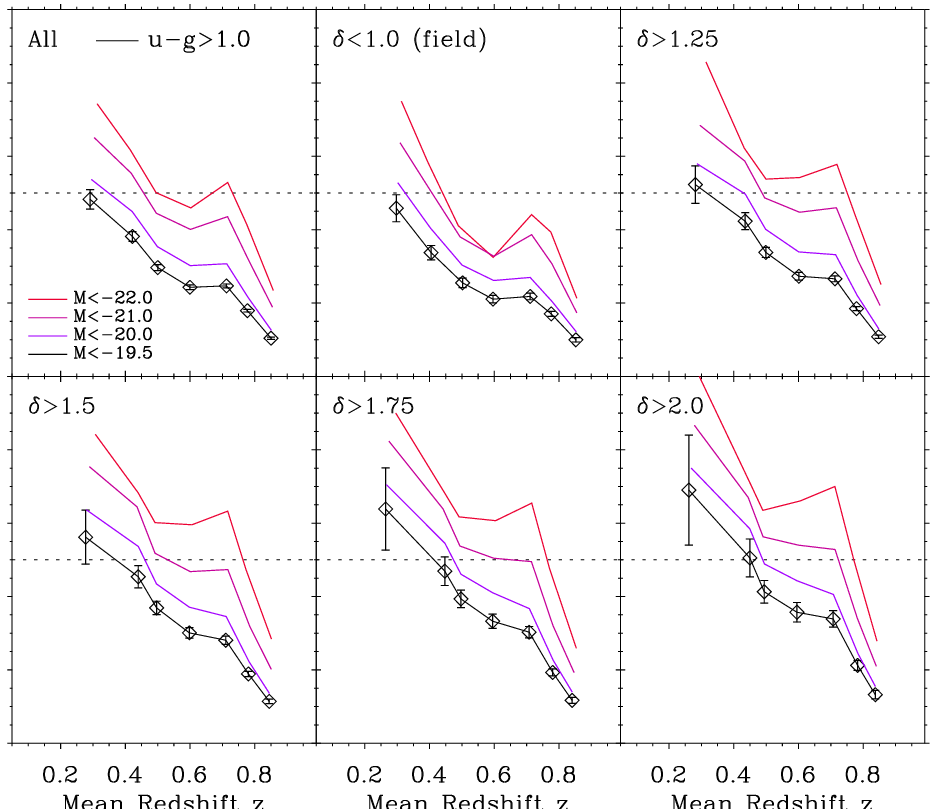}}
         }
    \caption{The observed evolution of galaxies out to $z\sim1$ in 
terms of galaxy number fractions for (a) morphology and (b) color as function 
of galaxy total luminosity $M_{g'}$ and the overdensity estimator $\delta_{15th}$.
Panel (a) shows the number fractions for $n > 2$ galaxies (S\'ersic 
index $n$). The crosses are the $n > 2$ number fractions based on SDSS galaxies. 
Panel (b) shows the number fractions for red galaxies (rest-frame 
$u^{*}-g'>1$). The fractions are calculated for seven redshift bins with 
increments of 0.1 and fixed widths of 0.2. The error bars are given by Poisson 
statistics only and are shown only for $M_{g'}<-19.5$ for clarity.}
    \label{fig:evol}
  \end{center}
\end{figure*}

We selected all galaxies with $r_{1/2}>$ 0\farcs1 and
$M_{g'}\leq-19.5$ mag (corresponding to $i{'}_{model}\sim24$ at
$z\sim1$) to keep only galaxies with reliable structural parameters
(see $\S3.2$). Due to low number statistics we adopt a lower redshift
limit of $0.2$. Furthermore we exclude $z>1$ objects due to their large 
uncertainties in $z_{phot}$. These cuts leave us with an 
essentially volume-limited sample of 28,911 galaxies with $0.2<z<1$ 
which we divide into $4$ equal redshift bins. Figure 
\ref{fig:red_seq} shows the $1/V_{acc}$ weighted rest-frame color-magnitude 
diagrams for the lowest ($0.2<z<0.4$ ) and highest ($0.8<z<1$) redshift 
bins as function of morphology and environment. Figure \ref{fig:evol} 
shows the redshift evolution of the number fractions of $n>2$ 
(``bulge-dominated'') galaxies and red (rest-frame\footnote[1]{In the 
remainder of this Letter we will use the $u^{*}-g'$ color always as being 
defined in the rest-frame of the galaxy.} $u^{*}-g'>1$) galaxies as a 
function of luminosity\footnote[2]{To avoid too small number statistics we 
take cumulative lower absolute magnitude limits instead of fixed absolute
magnitude bins.} and environment. We stress the quantitative nature of
our morphological classification scheme in that it measures the
concentration of the galaxy light profiles as parameterized by the
\sersic index $n$. We chose $n=2$ as our morphological discrimant
between ``bulge'' and ``disk'' dominated galaxies to be fully
consistent with \citet{hogg04} who studied similar 
morphology-color-magnitude relations for SDSS galaxies
($\bar{z}\sim0.1$). Their results are also plotted in Figure
\ref{fig:evol}a. We find good quantitative agreement (within the
errors) for our $M_{g'}<-20$ number fractions when extrapolated down
to the SDSS redshift of $z\sim0.1$. We summarize our main
results as follows:

1. Within the entire $M_{g'}<-19.5$ sample the $n > 2$ galaxy number fraction 
remains constant with redshift to within $\sim10\%$ (see Fig. \ref{fig:evol}a). 
At all redshifts, field populations have the lowest $n > 2$ fraction and this 
fraction remains constant with redshift. However, for overdense environments this 
fraction is larger and consistently increases towards lower redshifts, higher 
densities and higher luminosities showing that a strong morphological evolution 
has taken place since $z\sim1$. In overdense regions we observe that for $0.2<z<1$ 
the ``bulge-'' vs. ``disk-dominated'' galaxy ratio has increased by as much as a 
factor of $\sim2$ among bright $M_{g'}<-22$ populations relative to 
$M_{g'}<-19.5$ populations.  

2. At all redshifts, cluster-like environments also contain the highest fraction 
of red ($u^{*}-g'>1$) galaxies. With mostly blue ($u^{*}-g'<1$) galaxies at 
$z\sim1$, the red fraction increases towards lower redshifts and higher 
luminosities. The red fraction within cluster-like regions changes faster with 
redshift as compared to the field. We quantify this by fitting a line with slope 
$\beta$ to the observed color evolutions and find 
$\beta_{\delta>2}/\beta_{\delta<1}\sim1.6$ for $M_{g'}<-19.5$. 
We note the presence of a break at $z\sim0.7$ which marks a period of a less
strong ($\sim35\%$) color evolution down to $z=0.2$. 

The results discussed in 1 and 2 qualitatively agree with the postulations 
and predictions made by hierarchical formation models (e.g. Kauffmann \& Charlot 1998; 
Cole et al. 2000) where galaxies are initially formed as blue 
star-forming systems with $n<2$ or ``disk-dominated'' profiles. The observed trends 
show that the onset of a population's morphological evolution 
depends strongly on environment with the cluster-like environments exhibiting 
the strongest evolution. The latter would indicate that ``bulge-dominated'' 
galaxies are assembled at earlier times in denser environments which is in 
agreement with the predictions made by hierarchical merging models 
\citep[e.g.,][]{benson00,berlind03}. More specifically, they predict that 
mergers erase any galaxy disk signatures and lead to more spheroidal structures 
such as bulges \citep[e.g.,][]{barnes96} and that higher merger rates 
associated with denser environments \citep[e.g.,][]{lacey93} result in earlier 
onsets of morphological transformation. Mergers also lead to increased galaxy 
luminosity through the addition of stellar mass and ultimately result in more 
luminous and ``bulge-dominated'' populations. Our observations support this 
merging scenario.

3. In all environments, the color distribution of the observed galaxies is 
bimodal out to $z\sim1$ with a `gap' at $u^{*}-g'\sim1$
which is in agreement with the observations of \citet{bell04}. Rest-frame 
color-magnitude diagrams (Fig. \ref{fig:red_seq}) show that there exists 
a prominent red-sequence of galaxies at $0.2<z<0.4$ and a large blue-peak 
dominance at $0.8<z<1$. Equivalent color-magnitude diagrams for the redshift 
range $0.4<z<0.8$ (not included here) show bimodal distributions intermediate 
between these low and high redshift bins. The red peak ($u^{*}-g'>1$) of 
this bimodality appears to be greatly reduced towards higher redshifts and 
is predominantly being `built up' from ``bulge-dominated'' galaxies with 
decreasing redshift. This build-up proceeds more rapidly in more 
overdense environments. 

Based on their examination of the color bimodality, \citet{bell04} concluded 
that a combination of the interaction/merging of passively evolving galaxies 
and the truncation of their star formation would explain the main features of 
their observed color distribution. They suggest that the truncation may be 
associated with mergers that consume all the gas in an induced starburst. 
It is therefore interesting that we observe a strong color-magnitude relation 
(brighter being redder) at all redshifts and within all environments with a 
clear increase of the color evolutionary rate towards denser environments. 
The latter would agree well with an enhanced truncation of the star formation 
within denser environments. In addition to higher merger rates, denser 
environments also have other processes such as ram-pressure and that can 
affect galaxies travelling through their hot intergalactic medium 
\citep[e.g.,][]{quilis00, bekki02}.

The picture emerging from results presented here and elsewhere is one that 
supports basic tenets of hierarchical galaxy formation and evolution. The 
CFHTLS-Deep is an ongoing project, and the final dataset will allow us to 
quadruple our sample size with even greater sky coverage and depth. As is, 
our results show for the first time the redshift evolution of the 
color-magnitude-morphology relation over a single cosmological volume large 
enough to encompass many different environments ranging from the field to 
the most massive clusters.

\acknowledgments We would like to thank CFHT staff and members of the
CFHTLS community involved in the planning and execution of survey
observations and data processing. M. J. H. M. N. would especially like 
to thank the Herzberg Institute of Astrophysics (NRC-HIA) for their 
hospitality and financial support during the time when this work was conducted.

\clearpage

\clearpage

\end{document}